\documentstyle [12pt] {article}

\catcode`@=12
\begin{document}
\thispagestyle{empty}
\begin{flushright} UCRHEP-T209\\February 1998\
\end{flushright}
\vspace{0.5in}
\begin{center}
{\Large \bf Neutrino Masses and Leptogenesis\\
with Heavy Higgs Triplets\\}
\vspace{1.0in}
{\bf Ernest Ma\\}
\vspace{0.1in}
{\sl Department of Physics, University of California\\}
{\sl Riverside, California 92521, USA\\}
\vspace{0.5in}
{\bf Utpal Sarkar\\}
\vspace{0.1in}
{\sl Physical Research Laboratory\\}
{\sl Ahmedabad 380 009, India\\}
\vspace{1.0in}
\end{center}
\begin{abstract}\
A simple and economical extension of the minimal standard electroweak gauge 
model (without right-handed neutrinos) by the addition of two heavy Higgs 
scalar triplets would have two significant advantages.  \underline 
{Naturally} small Majorana neutrino masses would become possible, 
as well as leptogenesis in the early universe which gets converted at the 
electroweak phase transition into the present observed baryon asymmetry.
\end{abstract}

\newpage
\baselineskip 24pt

In the minimal standard electroweak gauge model, neutrinos are massless 
because there are no right-handed neutrino singlets and there is no Higgs 
scalar triplet.  The latter may be added\cite{1} so that the interaction
\begin{equation}
f_{ij} [\xi^0 \nu_i \nu_j + \xi^+ (\nu_i l_j + l_i \nu_j)/\sqrt 2 
+ \xi^{++} l_i l_j] + h.c.
\end{equation}
would induce a Majorana mass matrix of the neutrinos if $\xi^0$ has a small 
nonzero vacuum expectation value.  However, since the triplet $\xi$ carries 
lepton number, a massless Goldstone boson (the triplet Majoron) would appear if 
the model conserves lepton number before spontaneous symmetry breaking.  This 
would have counted as the equivalent of two extra neutrino flavors in the 
invisible decay of the $Z$ boson.  Hence it is already ruled out 
experimentally\cite{2}.  On the other hand, if lepton number is explicitly 
broken, this problem may not arise.  In particular, we will show in the 
following that if the scalar triplet is very heavy, {\it i.e.} of order 
$10^{13}$ GeV, then it is in fact natural for neutrinos to be of order 1 eV 
or less, and if there are two such triplets, their decays could generate a 
lepton asymmetry\cite{3} in the early universe which would get converted at 
the electroweak phase transition\cite{4} into the present observed baryon 
asymmetry.

Consider the most general Higgs potential of one doublet $\Phi = (\phi^+, 
\phi^0)$ and one triplet $\xi = (\xi^{++}, \xi^+, \xi^0)$:
\begin{eqnarray}
V &=& m^2 \Phi^\dagger \Phi + M^2 \xi^\dagger \xi \nonumber \\ 
&+& {1 \over 2} \lambda_1 (\Phi^\dagger \Phi)^2 + {1 \over 2} \lambda_2 
(\xi^\dagger \xi)^2 + \lambda_3 (\Phi^\dagger \Phi)(\xi^\dagger \xi) \nonumber 
\\ &+& \mu (\bar \xi^0 \phi^0 \phi^0 + \sqrt 2 \xi^- \phi^+ \phi^0 + \xi^{--} 
\phi^+ \phi^+) + h.c.
\end{eqnarray}
Let $\langle \phi^0 \rangle = v$ and $\langle \xi^0 \rangle = u$, then the 
conditions for the minimum of $V$ are given by
\begin{equation}
m^2 + \lambda_1 v^2 + \lambda_3 u^2 + 2 \mu u = 0,
\end{equation}
\begin{equation}
u (M^2 + \lambda_2 u^2 + \lambda_3 v^2) + \mu v^2 = 0.
\end{equation}
In the triplet Majoron model\cite{1}, $\mu = 0$, hence for $u \neq 0$, 
we have
\begin{equation}
v^2 = {{-\lambda_2 m^2 + \lambda_3 M^2} \over {\lambda_1 \lambda_2 - 
\lambda_3^2}}, ~~~ u^2 = {{\lambda_3 m^2 - \lambda_1 M^2} \over {\lambda_1 
\lambda_2 - \lambda_3^2}}.
\end{equation}
Since $v = 174$ GeV and $u$ should not be greater than a few keV (to suppress 
the $\gamma + e \rightarrow e$ + Majoron cross section\cite{5}), the parameter 
$M^2$ must be fine-tuned to equal $(\lambda_3/\lambda_1) m^2$ to extreme 
accuracy.  This model is thus rather unnatural, but of course it is also 
experimentally ruled out.  To see this, we consider the mass matrix spanned 
by the neutral scalar fields $\sqrt 2 Re \phi^0$ and $\sqrt 2 Re \xi^0$, 
{\it i.e.}
\begin{equation}
{\cal M}^2 = \left[ \begin{array} {c@{\quad}c} 2 \lambda_1 v^2 & 2 \lambda_3 
u v + 2 \mu v \\ 2 \lambda_3 u v + 2 \mu v & 2 \lambda_2 u^2 - \mu v^2/u 
\end{array} \right].
\end{equation}
If $\mu = 0$, then the eigenvalues of the above are $2 \lambda_1 v^2$ and 
$2 (\lambda_2 - \lambda_3^2/\lambda_1) u^2$.  The latter is the square of 
the mass of the partner of the Majoron and it is necessarily small.  Hence 
the $Z$ boson must decay into them if the model is correct.

If $\mu \neq 0$, then lepton number is explicitly violated and the mass 
matrix spanned by the neutral scalar fields $\sqrt 2 Im \phi^0$ and 
$\sqrt 2 Im \xi^0$ is given by
\begin{equation}
{\cal M}^2 = \left[ \begin{array} {c@{\quad}c} -4 \mu u & 2 \mu v \\ 2 \mu v 
& -\mu v^2/u \end{array} \right].
\end{equation}
The above contains one zero eigenvalue for the longitudinal component of the 
$Z$ boson, but the would-be Majoron is now massive with mass-squared given by
\begin{equation}
- {\mu \over u} (v^2 + 4 u^2) = (M^2 + \lambda_3 v^2) \left[ 1 + {\cal O} 
\left( {u^2 \over v^2} \right) \right],
\end{equation}
which is also approximately the mass-squared of its partner.  In the above, 
we have used Eq.~(4) and the fact that $u^2 << v^2$.  Hence 
$M^2 + \lambda_3 v^2$ must be positive and if it is also large enough, 
present experiments cannot rule out this model.

Let us in fact make $M$ very large.  In that case, we have a natural 
understanding of why $u$ can be so small because
\begin{equation}
u \simeq {{-\mu v^2} \over M^2},
\end{equation}
which is analogous to the usual seesaw mechanism for obtaining small 
Majorana neutrino masses, except that here we do not have any right-handed 
neutrinos.  [In a left-right gauge model, where there is already a 
right-handed neutrino, the left-handed neutrino also gets a mass from 
a Higgs triplet\cite{6} in addition to the canonical seesaw mechanism.] 
Substituting the above into Eq.~(3), we find
\begin{equation}
v^2 \simeq {{-m^2} \over {\lambda_1 - 2 \mu^2/M^2}}.
\end{equation}
Note that we have \underline {no fine tuning} ({\it i.e.} the cancellation 
of two large quantities to obtain a small one) in this model.

It may appear strange that the heavy triplet $\xi$ gets a tiny vacuum 
expectation value $u$.  However, this actually already occurs in the 
well-known case of the spontaneous breaking of $SU(5)$ using the {\bf 24} 
scalar representation.  Under $SU(3)_C \times SU(2)_L \times U(1)_Y$,
\begin{equation}
{\bf 24} = (1, 1, 0) + (8, 1, 0) + (1, 3, 0) + (3, 2, -5/6) + (3^*, 2, 5/6).
\end{equation}
What everyone knows is that a large vacuum expectation value $v_1$ for the 
$(1, 1, 0)$ component breaks $SU(5)$ down to the standard-model gauge group. 
What many people do not realize is that the further breaking of $SU(2)_L 
\times U(1)_Y$ down to $U(1)_Q$ using the fundamental {\bf 5} scalar 
representation necessarily induces a small vacuum expectation value $v_3$ for 
the heavy $(1, 3, 0)$ component.  It has been shown recently\cite{7} that 
$v_3 \sim v_2^2 / v_1$, where $v_2$ is the electroweak vacuum expectation 
value.  Again the seesaw structure appears automatically.

Another way of handling the heavy Higgs triplet is to integrate it out.  From 
Eqs.~(1) and (2), we obtain the effective nonrenormalizable term
\begin{equation}
{{-f_{ij} \mu} \over {M^2}} [\phi^0 \phi^0 \nu_i \nu_j - \phi^+ \phi^0 
(\nu_i l_j + l_i \nu_j) + \phi^+ \phi^+ l_i l_j] + h.c.
\end{equation}
>From Eq.~(2) itself, the reduced Higgs potential involving only $\Phi$ is
\begin{equation}
V = m^2 \Phi^\dagger \Phi + {1 \over 2} \left( \lambda_1 - {{2\mu^2} \over 
M^2} \right) (\Phi^\dagger \Phi)^2,
\end{equation}
the last term coming\cite{7} from the exchange of $\xi$.  As $\phi^0$ acquires 
a nonzero vacuum expectation value $v$, we obtain Eq.~(10) as we should, and 
the neutrino mass matrix becomes $-2 f_{ij} \mu v^2 / M^2 = 2 f_{ij} u$ as 
expected.

Armed with Eq.~(9) and making the reasonable assumption that $|\mu|$ is 
of order $M$ or less, we find, for $M \sim 10^{13}$ GeV, that $u$ is less 
than a few eV.  This is then a suitable natural mechanism for small 
Majorana neutrino masses.  Furthermore, a mass of $10^{13}$ GeV or so is very 
evocative of the natural energy scale for leptogenesis\cite{3}.  For this, 
we would need (at least) two such heavy Higgs triplets to have the proper 
CP violation which distinguishes states of different lepton number.

We now write down the mass terms and the Yukawa couplings of the heavy Higgs 
triplets ($\xi_a$, $a = 1,2$), which are relevant for the study of 
leptogenesis in this scenario: 
\begin{eqnarray}
-{\cal L} &=& \sum_{a=1,2} \left\{ M_a^2 \xi_a^\dagger \xi_a + \left( 
f_{aij} [\xi_a^0 \nu_i \nu_j + \xi^+_a (\nu_i l_j + 
l_i \nu_j)/\sqrt 2 + \xi^{++}_a l_i l_j] \right. \right. \nonumber \\
&&\left. \left. + \mu_a [\bar \xi_a^0 \phi^0 \phi^0 + \sqrt{2} \xi^-_a
\phi^+ \phi^0 + \xi_a^{--} \phi^+ \phi^+] + h.c. \right) \right\}. 
\end{eqnarray}
At an energy scale far above that of electroweak symmetry breaking, 
$SU(2)_L \times U(1)_Y$ gauge invariance means that we can pick any one of 
the three components of the triplet for consideration and the results are 
guaranteed to be valid for the other two.  Let us choose $\xi_a^{++}$, then 
their decays are:
\begin{equation}
\xi_a^{++} \rightarrow \left\{ \begin{array} {l@{\quad}l} l_i^+ l_j^+ & 
(L = -2) \\ \phi^+ \phi^+ & (L = 0) \end{array} \right.
\end{equation}
The coexistence of the above two types of final states indicates the 
nonconservation of lepton number.  On the other hand, any lepton asymmetry 
generated by $\xi_a^{++}$ would be neutralized by the decays of $\xi_a^{--}$, 
unless CP conservation is also violated and the decays are out of thermal 
equilibrium\cite{8} in the early universe.

We will use the effective mass-matrix formalism\cite{9} to discuss the 
generation of lepton asymmetry in this model.  We note that in the often 
studied case of the decays of heavy singlet neutrinos, there is always 
$\nu_R - \bar \nu_R$ mixing, whereas $\xi_a^0 - \bar \xi_b^0$ mixing is 
strictly forbidden here before $SU(2)_L \times U(1)_Y$ symmetry breaking.  
Without loss of generality, we can choose the tree-level mass matrix for the 
triplets $\xi_a$ to be diagonal and real, as already assumed in Eq.~(14).  
Hence CP is conserved at this level.  However, CP nonconservation may still 
occur at the one-loop level due to the interference between tree and one-loop 
diagrams, as shown in Fig.~1.  
\begin{figure}[t]
\vskip 2.5in\relax\noindent\hskip -.5in\relax{\includegraphics{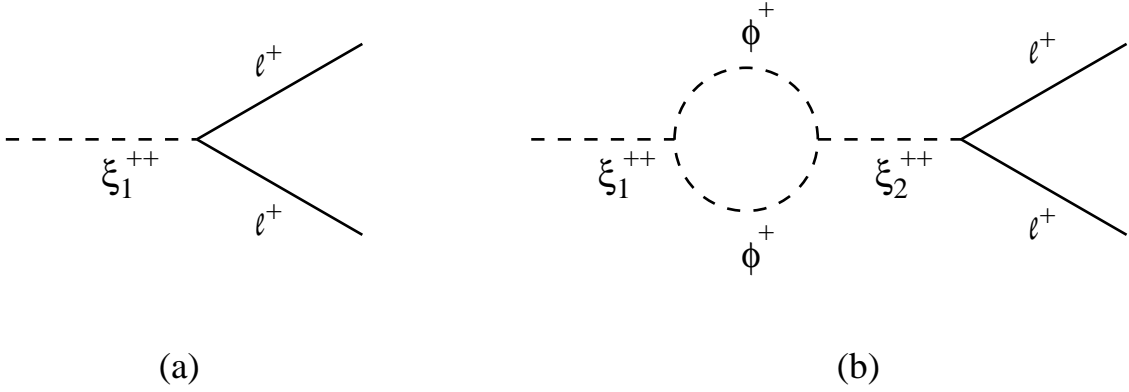}}
\caption{The decay of $\xi_1^{++} \to l^+ l^+$ at tree level (a) and in 
one-loop order (b).  A lepton asymmetry is generated by their interference.}
\end{figure}
We note also that in all other 
previous models of baryogenesis (or leptogenesis), where the decays of heavy 
particles generate the asymmetry, there is always a one-loop vertex 
correction, whereas in this model, there is none.  However, leptogenesis is 
still possible if there are two triplets because off-diagonal one-loop 
self-energy terms will have absorptive parts. 

Before we present the details of our calculation, let us consider how CP 
nonconservation appears in Eq.~(14).  If there is only one $\xi$, then the 
relative phase between any $f_{ij}$ and $\mu$ can be chosen real.  Hence 
a lepton asymmetry cannot be generated.  With two $\xi$'s, even if there is 
only one lepton family, one relative phase must remain among the quantities 
$f_1$, $f_2$, $\mu_1$, and $\mu_2$.  As for the possible relative phases 
among the $f_{aij}$'s, they cannot generate a lepton asymmetry because they 
all refer to final states of the same lepton number.

In the presence of interactions, the diagonal tree-level mass matrix 
$M_a^2$ is replaced by
\begin{equation}
{1 \over 2} \xi_a^\dagger ({\cal M}^2_+)_{ab} \xi_b + {1 \over 2} 
(\xi_a^*)^\dagger ({\cal M}^2_-)_{ab} \xi_b^*,
\end{equation}
where
\begin{equation}
{\cal M}^2_\pm = \left( \begin{array} {c@{\quad}c} M_1^2 -i {\cal G}_{11} & 
-i {\cal G}_{12}^\pm \\ -i {\cal G}_{21}^\pm & M_2^2 -i {\cal G}_{22} 
\end{array} \right),
\end{equation}
where ${\cal G}_{ab}^+ = \Gamma_{ab} M_b$, ${\cal G}_{ab}^- = \Gamma_{ab}^* 
M_b$, and ${\cal G}_{aa} = \Gamma_{aa} M_a$.  Now
\begin{equation}
\Gamma_{ab} M_b = {1 \over 8 \pi} \left( \mu_a \mu_b^* + M_a M_b \sum_{k,l} 
f^*_{akl} f_{bkl} \right),
\end{equation}
and it comes from the absorptive part of the one-loop self-energy diagram for 
$\xi_b \to \xi_a$ which is of course equal to that for $\xi_a^* \to \xi_b^*$. 
Hence $\Gamma_{12} M_2 = \Gamma^*_{21} M_1$ as expected.  If there is no phase 
convention which allows us to choose $\Gamma_{ab}$ to be real, then CP 
conservation is violated.

Assuming that $\Gamma_a \equiv \Gamma_{aa} << M_a$, we solve for the 
eigenvalues of ${\cal M}^2_\pm$:
\begin{equation}
\lambda_{1,2} = {1 \over 2} (M_1^2 + M_2^2 \pm \sqrt{\cal S}),
\end{equation}
where $ {\cal S} = (M_1^2 - M_2^2)^2 - 4 \left| \Gamma_{12} M_2 \right|^2,
$ and $M_1 > M_2$ is assumed.  The corresponding physical states are
\begin{equation}
\psi_{1,2}^+ = a_{1,2}^+ {\xi_1 } + b_{1,2}^+ {\xi_2 }, ~~~ 
\psi_{1,2}^- = a_{1,2}^- \xi_1^* + b_{1,2}^- \xi_2^*,
\end{equation}
where $a_1^\pm = b_2^\pm = 1/{\cal N}$, $b_1^\pm = {\cal C}_1^\pm /{\cal N}$, 
$a_2^\pm = {\cal C}_2^\pm / {\cal N}$, with
\begin{equation}
{\cal C}_1^+ = -{\cal C}_2^- = \frac{-2i\Gamma_{12}^* M_2}{M_1^2 - M_2^2 + 
\sqrt{{\cal S}}}, \hskip .2in {\cal C}_1^- = -{\cal C}_2^+ = 
\frac{-2i\Gamma_{12} M_2}{M_1^2 - M_2^2 + \sqrt{{\cal S}}},
\end{equation}
and ${\cal N} = \sqrt{1+|{\cal C}_i^\pm|^2}$.  Note that whereas $\xi_a$ and 
$\xi_a^*$ are CP conjugates, $\psi_i^+$ and $\psi_i^-$ are not, even though 
they have the same mass.  This is analogous to having physical Majorana 
neutrinos which are not CP eigenstates, as discussed in Ref.[9].

The physical states $\psi_{1,2}^\pm$ will now evolve with time and decay 
into leptons and antileptons.  The resulting lepton asymmetries $\delta_i = 
n_l/n_i$ will be given by
\begin{equation}
\delta_i =  2  \left[ B \left( \psi_i^- 
\to l l \right) - B \left( \psi_i^+ \to l^c l^c \right) \right],
\end{equation}
where $n_i$ is the number density of $\psi_i^\pm$, and
\begin{equation}
B \left( \psi_i^- \to l l \right) = {{\sum_{k,l} \left| a_i^- f^\ast_{1kl} + 
b_i^- f^\ast_{2kl}  \right|^2} \over {\left| a_i^- \mu_1 + b_i^- \mu_2 
\right|^2 / M_i^2 + \sum_{k,l} \left| a_i^- f^*_{1kl} + b_i^- f^*_{2kl} 
\right|^2}},
\end{equation}
\begin{equation}
B \left( \psi_i^+ \to l^c l^c \right) = {{\sum_{k,l} \left| a_i^+ f_{1kl} + 
b_i^+ f_{2kl} \right|^2} \over {\left| a_i^+ \mu_1^* + b_i^+ \mu_2^* 
\right|^2 / M_i^2 + \sum_{k,l} \left| a_i^+ f_{1kl} + b_i^+ f_{2kl} 
\right|^2}}.
\end{equation}
Assuming $(M_1^2 - M_2^2)^2 \gg 4|\Gamma_{12} M_2|^2$, so that 
$\sqrt {\cal S} \simeq M_1^2 - M_2^2$, we get
\begin{equation}
\delta_i \simeq 
{{Im \left[ \mu_1 \mu_2^* \sum_{k,l} f_{1kl} f_{2kl}^* \right]} \over 
{8 \pi^2 (M_1^2 - M_2^2)}} \left[ {{ M_i} \over 
\Gamma_i}  \right].
\end{equation}
Note that there is no contribution from the purely leptonic term
because it is identically zero as expected.

In calculating the lepton asymmetry $\delta_i$, we have assumed that when the 
temperature was much higher than the masses of the $\psi$'s, there was no 
lepton asymmetry.  Only around the time when the $\psi$'s started decaying was 
a lepton asymmetry created.  At that time, these scalars also became 
nonrelativistic.  In the case $M_1 > M_2$ as we have assumed, when the universe 
cooled down to below $M_1$, most of $\psi_1$ would decay away.  However, the 
asymmetry so created would be erased by the lepton-number nonconserving 
interactions of $\psi_2$.  Hence only the subsequent decay of $\psi_2$ 
at $T < M_2$ would generate a lepton asymmetry to remain until the onset of 
the electroweak phase transition.  This asymmetry would evolve with time 
following the Boltzmann equation,
\begin{equation}
\frac{{\rm d}n_l}{{\rm d}t} + 3 H n_l = \delta_2
\Gamma_2 [n_2 - n_2^{eq} ] - \left(
\frac{n_l}{n_\gamma} \right) n_2^{eq} \Gamma_2
-2 n_\gamma n_l \langle \sigma |v| \rangle
\end{equation}
The second term on the left side comes from the expansion of the universe,  
where $H = 1.66 g_*^{1/2} (T^2 / M_{Pl})$ is the Hubble constant with $g_*$  
the effective number of massless particles.  $\Gamma_2$ is the  thermally  
averaged  decay rate of $\psi_2$,  $n_\gamma$ is the photon  density and the
term  $\langle  \sigma  |v|  \rangle$   describes  the  thermally
averaged cross section of $l + l \leftrightarrow \phi +
\phi$  scattering.  The density of $\psi_2$ satisfies the
Boltzmann equation,
\begin{equation}
\frac{{\rm d}n_2}{{\rm d}t} + 3 H n_2 = - 
\Gamma_2 (n_2 - n_2^{eq})
\end{equation}

It is now convenient to use the dimensionless variable $x  =
{M_2}/{T}$ as well as the particle density per entropy density $Y_i
= {n_i}/{s}$, and the relation $t = {x^2}/{ 2H (x = 1) }$.  We also define 
the parameter $K \equiv \Gamma_2 (x = 1) / H(x = 1)$ 
as a measure of the deviation from equilibrium. For $K << 1$ at 
$T \sim M_2$, the system is far from equilibrium; hence the last two terms 
responsible for the depletion of $n_l$ would be negligible.  With these 
simplifications and the above redefinitions, the Boltzmann equations 
effectively read:
\begin{equation}
\frac{{\rm d}Y_l}{{\rm d}x} = (Y_2 - Y_2^{eq}) \delta_2 K x, ~~~ 
\frac{{\rm d}Y_2}{{\rm d}x} = - (Y_2 - Y_2^{eq}) K x.  
\end{equation}
In this limit $K << 1$, it is not difficult to obtain an 
asymptotic solution for $n_l$. Although the decay rate of 
$\psi_2$ is not fast enough to bring the number density $n_2$ to 
its equilibrium density, it is a good approximation to assume 
that the universe never goes far away from equilibrium. In 
other words, we can assume  
${\rm d}(Y_2 - Y_2^{eq})/{\rm d}x = 0$ to 
get an asymptotic value for $Y_l$, given by $Y_l = n_l/s = \delta_2 / g_*$. 
However, if $K > 1$, the terms which deplete $n_l$ 
dominate for some time and the lepton number density  
reaches its new asymptotic value, which is lower than the value 
it reaches in the out-of-equilibrium case. 
In this case although it is difficult to get 
an analytic solution of the Boltzmann equations, it is possible\cite{10} 
to get an approximate suppression factor given by 
$1 / K ({\rm ln} K)^{0.6}$.

The lepton asymmetry thus generated after the Higgs triplets 
decayed away would be the same as the $(B-L)$ asymemtry before the
electroweak phase transition. During the electroweak phase 
transition, the presence of sphaleron fields would relate
this $(B-L)$ asymmetry to the baryon asymmetry of the universe\cite{11}. The
final baryon asymmetry thus generated can then be given by the
approximate relation 
\begin{equation}
{n_B \over s} \sim {\delta_2 \over 3 g_* K ({\rm ln} K)^{0.6}}
\end{equation}

It is clear from Eq.~(25) that we must have two Higgs triplets for $n_l$ to 
be nonzero.  Hence the neutrino mass matrix is now given by
$(m_{\nu})_{ij} = -2 v^2 ( f_{1ij} \mu_1 / M_1^2 + f_{2ij} \mu_2 / M_2^2)$. 
Similarly, the effective quartic coupling of the Higgs doublet $\Phi$ 
is now modified in Eqs.~(10) and (13) to read $\lambda_1 - 2\mu_1^2/M_1^2 
- 2\mu_2^2/M_2^2$.  This means that we have the flexibility to choose $M_1$ 
and $M_2$ somewhat differently but within an order of magnitude to obtain a 
neutrino mass of order 1 eV, as well as the observed baryon 
asymmetry of the universe.  For example, let $M_2 = 10^{13}$ GeV, $\mu_2 = 
2 \times 10^{12}$ GeV, and $f_{233} = 1$, then $m_{\nu_\tau} = 1.2$ eV, 
assuming that the $M_1$ contribution is negligible.  Now let $M_1 = 3 \times 
10^{13}$ GeV, $\mu_1 = 10^{13}$ GeV, and $f_{1kl} \sim 0.1$, then the decay of 
$\psi_2^\pm$ generates a lepton asymmetry $\delta_2$ of about $8 \times 
10^{-4}$ if the CP phase is maximum.  Using $M_{Pl} \sim 10^{19}$ GeV and $g_* 
\sim 10^2$, we find $K \sim 2.4 \times 10^{3}$.  Hence $n_B/s \sim 10^{-10}$ 
as desired.

In conclusion, we have presented in this paper a simple and economical 
extension of the minimal standard model to obtain naturally small Majorana 
neutrino masses and explain the observed baryon asymmetry of the universe.  
This is achieved by the addition of two heavy Higgs triplets.  We show that it 
is in fact natural for them to have very small nonzero vacuum expectation 
values.  For neutrino masses of less than a few eV, the mass scale of these 
triplets is of order $10^{13}$ GeV, which is very suitable for 
leptogenesis.  We then calculate the lepton asymmetry, using a newly developed 
effective mass-matrix formalism\cite{9}.  Our proposal is an equally viable and 
attractive alternative to the canonical scenario where there are three 
additional right-handed singlet neutrinos with large Majorana masses.  They 
are distinguished in principle by the fact that the seesaw mechanism in the 
latter case decreases very slightly the coupling of the left-handed neutrinos, 
whereas there is no such deviation at all in our case.  Otherwise, they are 
identical in their two significant advantages over the minimal standard model, 
{\it i.e.} naturally small neutrino masses and leptogenesis.

Note added: Equation (2) is missing the term $\Phi^\dagger_i \tau^a_{ij} 
\Phi_j \xi^\dagger_b t^a_{bc} \xi_c$.  Hence $\lambda_3$ in all subsequent 
equations should include the coupling of this term as well.

\vskip 0.5in
\begin{center} {ACKNOWLEDGEMENT}
\end{center}

E.M. thanks the Physical Research Laboratory, Ahmedabad for hospitality 
during a short visit.  This work was supported in part by the U.~S.~Department 
of Energy under Grant No.~DE-FG03-94ER40837.

\newpage
\bibliographystyle{unsrt}

\end{document}